\documentclass[aps,reprint, prx, superscriptaddress, showpacs,nobibnotes]{revtex4-2}
\usepackage{amsmath}
\usepackage{graphicx}
\usepackage[colorlinks=true, allcolors=blue]{hyperref}
\usepackage{appendix}
\usepackage{dcolumn}
\usepackage[version=4]{mhchem}

\begin{document}
\title{Measurement of the magnetic octupole susceptibility of \ce{PrV2Al20}}
\author{Linda Ye}
\thanks{These authors contributed equally}
\altaffiliation{Present Address: Division of Physics, Mathematics and Astronomy, California Institute of Technology, Pasadena, CA 92115, USA}
\email{lindaye0@stanford.edu}
\affiliation{Department of Applied Physics, Stanford University, Stanford, CA 94305, USA}
\affiliation{Geballe Laboratory for Advanced Materials, Stanford University, California 94305, USA}
\author{Matthew E. Sorensen}
\thanks{These authors contributed equally}
\affiliation{Geballe Laboratory for Advanced Materials, Stanford University, California 94305, USA}
\affiliation{Department of Physics, Stanford University, Stanford, CA 94305, USA}
\author{Maja D. Bachmann}
\affiliation{Department of Applied Physics, Stanford University, Stanford, CA 94305, USA}
\affiliation{Geballe Laboratory for Advanced Materials, Stanford University, California 94305, USA}
\author{Ian R. Fisher}
\email{irfisher@stanford.edu}
\affiliation{Department of Applied Physics, Stanford University, Stanford, CA 94305, USA}
\affiliation{Geballe Laboratory for Advanced Materials, Stanford University, California 94305, USA}
\maketitle

\textbf{
In the electromagnetic multipole expansion, magnetic octupoles are the subsequent order of magnetic multipoles allowed in centrosymmetric systems, following the more commonly observed magnetic dipoles. As order parameters in condensed matter systems, magnetic octupoles have been experimentally elusive. In particular, the lack of simple external fields that directly couple to them makes their experimental detection challenging. Here, we demonstrate a methodology for probing the magnetic octupole susceptibility using a product of magnetic field $H_i$ and shear strain $\epsilon_{jk}$ to couple to the octupolar fluctuations, while using an adiabatic elastocaloric effect to probe the response to this composite effective field. We observe a Curie-Weiss behavior in the obtained octupolar susceptibility of \ce{PrV2Al20} up to temperatures approximately forty times the putative octupole ordering temperature. Our results demonstrate the presence of magnetic octupole fluctuations in the particular material system, and more broadly highlight how anisotropic strain can be combined with magnetic fields to formulate a versatile probe to observe otherwise elusive emergent `hidden' electronic orders.
}

\begin{figure*}[t]
\includegraphics[width =1.9 \columnwidth]{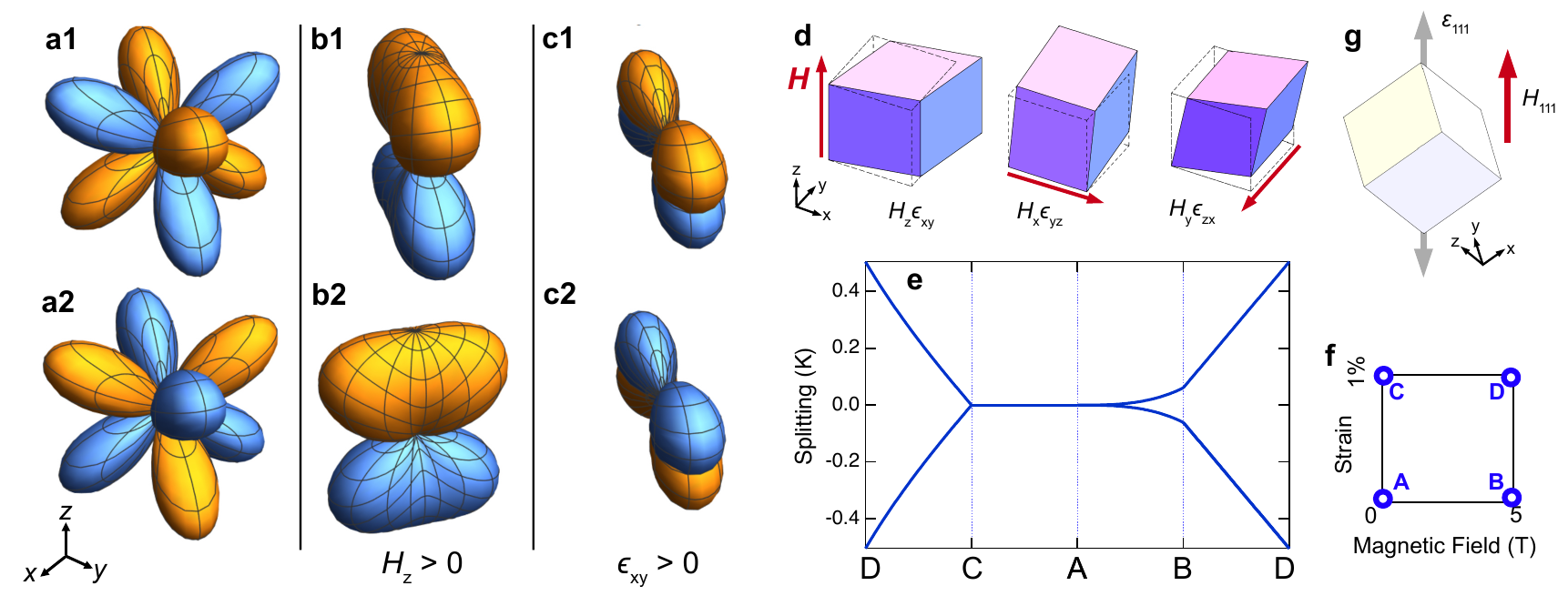}
\caption{\label{fullIntuitive} \textbf{Interplay between $\overline{J_xJ_yJ_z}$ magnetic octupole with the composite strain/magnetic field} (a-c) Illustration of the magnetic charge density of the positive and negative $\overline{J_xJ_yJ_z}$ octupolar moments in the CEF groundstate of PrV$_2$Al$_{20}$, shown via contours of constant magnetic charge (blue and orange lobes indicate positive and negative sign of magnetic charge, respectively). Panel (a1,a2) shows the case for $\epsilon = H = 0$; Panel (b1,b2) illustrates the effect of a $z$-axis magnetic field $H_z$; and panel (c1,c2) illustrates the effect of a $\Gamma_5$ strain $\epsilon_{xy}$. The magnetic charge density distribution shown in (a1) and (a2) are time-reversal pairs. (d) Schematic of the three components of $H_i\epsilon_{jk}$ with cuboids illustrating the lattice deformations with respect to the undeformed lattice (dashed lines) and red arrows the perpendicular magnetic field. (e,f) Energetic splitting of the ground state doublet (e, see text and Supplementary Materials) for selected linear cuts in the 2D space of applied strain and magnetic field (f). (g) Illustration of the orientation of applied uniaxial stress and magnetic field (both along [111]) in our experiment with respect to the principal cubic axes. The strain controlled in the experiments $\varepsilon_{111}\equiv\Delta L_{111}/L_{111}$ is also indicated. }\label{fig-1}
\end{figure*}

Within the Landau paradigm, generic electronic and magnetic ordered states are characterized by symmetry-breaking order parameters, and the electromagnetic multipole expansion provides a powerful framework to further characterize these. Ordered states based on electric and magnetic dipoles are common, and the corresponding order parameters can be readily coupled to, and manipulated by, uniform electric and magnetic fields respectively.  In centrosymmetric systems, the magnetic octupole is the next-allowed magnetic multipolar degree of freedom, but is much less commonly encountered. In particular, interactions between higher rank multipoles fall off
more rapidly than the simpler dipole case, making such states very rare. Initially discussed as a `hidden' degree of freedom based on localized electron orbitals \cite{Santini2009-tn,Matsumura2017-tm,Sibille2020-qd,PhysRevLett.124.087206}, magnetic octupoles have more recently witnessed a growing application as a notion in describing topological antiferromagnets with non-collinear spin structures \cite{Suzuki2017-oz,Wang2017-om,Higo2018-oq,nakatsuji2022topological}, and spin polarization in the electronic structure of altermagnets \cite{vsmejkal2022emerging,bhowal2022magnetic}. The difficulty of generating the most natural conjugate field for these octupolar moments --\textit{i.e.} microscopic magnetic fields with octupolar angular distribution-- makes the experimental detection and control of such orders and associated fluctuations generally challenging, motivating the development of novel experimental methods to couple to and probe these elusive degrees of freedom.

In this work we experimentally demonstrate a thermodynamic probe of magnetic octupolar fluctuations utilizing a combination of strain and magnetic field which (acting together as a composite effective conjugate field) couples bilinearly to magnetic octupoles. In particular, we focus on measuring the octupolar susceptibility, defined analoguously to the magnetic susceptibility as the rate of change of the induced octupolar moment in response to the conjugate field. In contrast to measuring the spontaneous octupole moment below a phase transition, measuring the susceptibility for temperatures above a phase transition is especially useful since this is a quantity (a) that is finite for all temperatures, (b) the divergence of which directly attests to the presence of growing fluctuations, and (c) which can be directly compared to other competing symmetry channels (\textit{i.e.} even in cases where a lower rank multipole ‘wins’, identifying the presence of strong fluctuations of the competing higher rank multipole state can help determine pathways to realizing such a state in other related materials). Here we experimentally identify strong octupolar fluctuations in the cubic compound \ce{PrV2Al20}, by measuring the temperature-dependence of the octupolar susceptibility.

\ce{PrV2Al20} belongs to a family of praseodymium-based cubic cage compounds Pr$T_2X_{20}$ ($T$= Ir, Rh, $X$= Zn; $T$= V, Ti, $X$= Al) \cite{Onimaru2010-bm,Sakai2011-ks,Sakai2012-in,Tsujimoto2014-cm,Onimaru2011-or,VHeatCapacity,Onimaru2016-jz}. In Pr$T_2X_{20}$ the $J=4$ $4f^2$ state of \ce{Pr^{3+}} takes a doublet crystal field ground state in the local $T_d$ environment with $\Gamma_3$ symmetry \cite{Onimaru2016-jz}. The $\Gamma_3$ non-Kramers doublet \cite{Lea1962-fp} is non-magnetic, such that dipolar moments are forbidden within the manifold ( $\langle \Gamma_3^a|J_i|\Gamma_3^b\rangle=0$, $i=x,y,z$, $a,b=1,2$); this makes the family of compounds an ideal platform to study higher-rank multipolar order and fluctuations, as well as their interaction with conduction electrons \cite{Onimaru2010-bm,Sakai2011-ks,Sakai2012-in,Tsujimoto2014-cm,Onimaru2011-or,VHeatCapacity,Onimaru2016-jz,Freyer2018-nf,Lee2018-cn}. The allowed order parameters within the manifold are two distinct electric quadrupoles ($3J_z^2-J^2$, $J_x^2-J_y^2$) and a magnetic octupole ($\overline{J_xJ_yJ_z}$, equivalent to the notation $\mathcal{T}_{xyz}$ \cite{Suzuki2017-oz}); the former (latter) belong to $\Gamma_3$ ($\Gamma_2$) irreducible representations of $T_d$. We have specifically chosen \ce{PrV2Al20} to probe octupolar fluctuations because of its well-isolated crystal electric field (CEF) $\Gamma_3$ ground state ($40$ K between ground and first excited states \cite{Sakai2011-ks}). The material undergoes a succession of two closely-spaced continuous phase transitions at low temperatures. While the exact character of these phases is unknown, the first of these (at $0.75$ K) is thought to be to an antiferroquadrupolar state \cite{Sakai2011-ks} while it has been suggested that the second phase transition at $0.65$ K might have an octupolar character \cite{VHeatCapacity}. 

The spatial distribution of the magnetic charge density of the $\overline{J_xJ_yJ_z}$ octupolar moments is depicted in Fig. \ref{fig-1}(a1,a2) with (a1) and (a2) forming a time-reversal pair. Due to their complex angular distribution, uniform magnetic and strain fields (which linearly couple to lower-order magnetic dipoles and electrical quadrupoles, respectively) are not able to distinguish the positive and negative octupole moments. However, two higher-order quantities are allowed by symmetry to couple bilinearly to the octupole moments \cite{YongBaek,SelfCiteGarbage}: a third-order magnetic field tensor $H_iH_jH_k$, and alternatively, a combination of strain and magnetic field of the form $H_i\epsilon_{jk}$ (where $i,j,k \in {x,y,z}$ and $i\neq j \neq k$). The latter field, $H_i\epsilon_{jk}$ requires a magnetic field along one of the principal cubic axes ($H_i$) and a shear strain in the plane perpendicular to that axis ($\epsilon_{jk}$); although neither $H_i$ nor $\epsilon_{jk}$ alone splits the $\Gamma_3$ doublet to first order, two equivalent second order interactions of the octupolar moment with the field (strain) and then strain (field) provide the mechanism by which the combination acts as an effective field. Starting from the positive or negative octupolar moment (Fig. \ref{fig-1}(a1,a2)), a vertical magnetic field $H_z$ deforms the magnetic charge density of the octupolar moments via a Zeeman interaction and induces a component with opposite sign of the electric quadrupole $J_xJ_y+J_yJ_x$ for the two octupoles (Fig. \ref{fig-1}(b1,b2)) \cite{Matsumura2017-tm}; the induced quadrupole can then couple to a shear strain $\epsilon_{xy}$ via magnetoelastic coupling \cite{luthi2007physical}, differentiating the two octupoles. The same steps can be applied in reverse: a shear strain $\epsilon_{xy}$ induces a magnetic dipole moment of opposite sign depending on the octupole moment (Fig. \ref{fig-1}(c1,c2)), which can then be split by $H_z$. Since the magnetic octupole, magnetic field and shear strains belong to $\Gamma_2,\Gamma_4$ and $\Gamma_5$ irreducible representations of $T_d$ point group respectively, the two interaction paths illustrated above can be expressed as $\Gamma_4 \otimes \Gamma_2 = \Gamma_5$ and $\Gamma_5 \otimes \Gamma_2 = \Gamma_4$, respectively. There are in total three components $H_x\epsilon_{yz},H_y\epsilon_{zx}$ and $H_z\epsilon_{xy}$ illustrated in Fig. \ref{fig-1}(d) that constitute the composite, effective `octupolar' field.

At the microscopic level, the coupling of $H_i\epsilon_{jk}$ with the octupolar degrees of freedom within the $\Gamma_3$ doublet is based on a second-order virtual process involving excited crystal field states, while that of $H_iH_jH_k$ invokes third order perturbations (see Supplementary Materials). In Fig. \ref{fig-1}(e) we illustrate the anticipated splitting within the ground state manifold induced by the presence of strain and/or magnetic field, along paths in the field-strain phase space depicted in Fig. \ref{fig-1}(f). In the presence of only strain (path A-C), the splitting is negligible (the only expected contribution arises from admixing of $\Gamma_5$ quadrupole states from the excited CEF levels), while in the presence of only magnetic field (path A-B), a $H^3$ splitting from third order processes is expected; the splitting is significantly enhanced in the presence of both strain and magnetic fields (paths C-D and B-D).

\begin{figure}
    \includegraphics[width = \columnwidth]{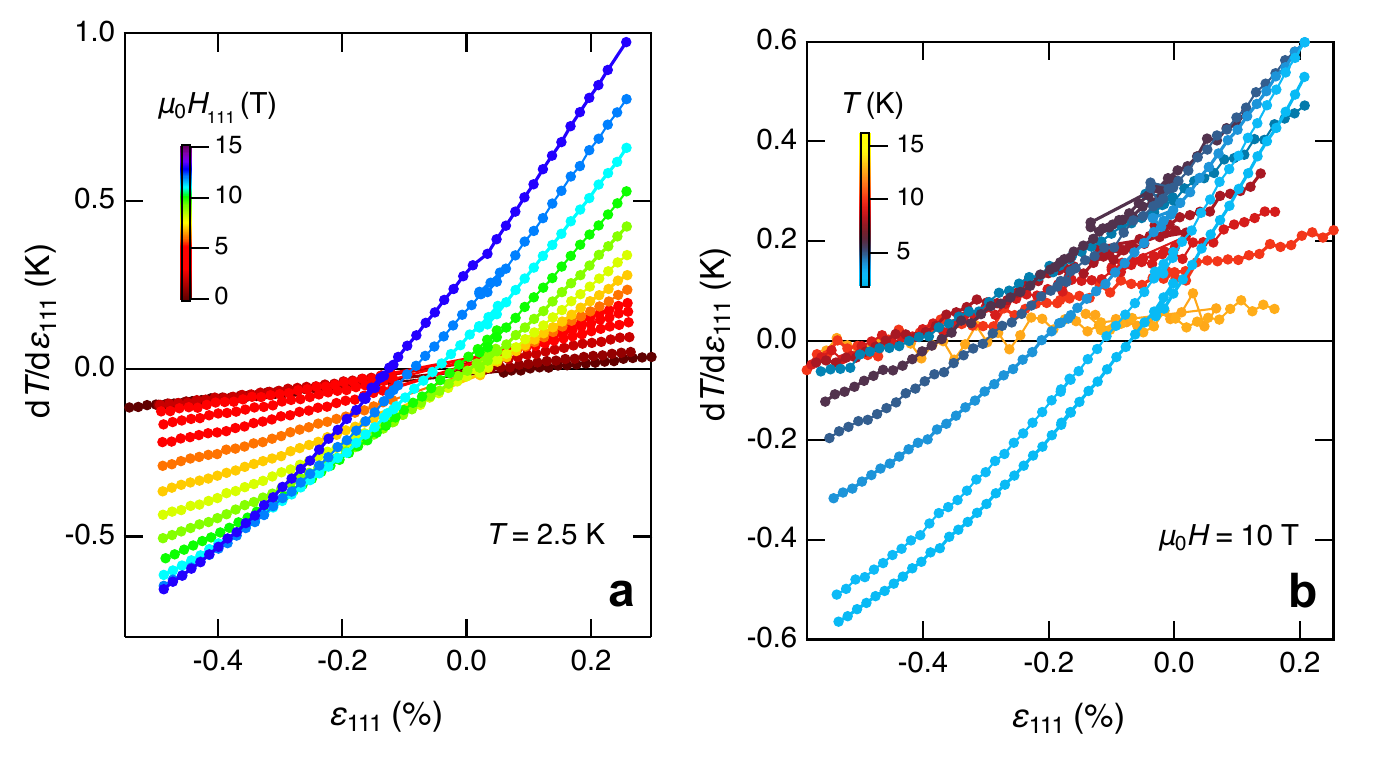}
    \caption{\textbf{Temperature- and field-dependence of the elastocaloric coefficient in \ce{PrV2Al20}} The elastocaloric coefficient $dT/d\varepsilon_{111}$ as a function of the DC longitudinal strain in the [111] direction, $\varepsilon_{111}$ (a) for a constant temperature $T=2.5$ K and several representative magnetic fields; and (b) at a constant magnetic field $\mu_0H=10$ T and various representative temperatures. The color scale indicating the respective magnetic field and temperature for each curve is inset. }
    \label{fig-2}
\end{figure}

Experimentally, we opted for a configuration with both magnetic field and uniaxial stress oriented along [111] of the single crystal (Fig. \ref{fig-1}(g), more see Supplementary Materials). This configuration maximizes $H_i\epsilon_{jk}$ (invoking all three permutations) and avoids strain or high-order magnetic field components that couple bilinearly to the allowed quadrupoles in the $\Gamma_3$ CEF manifold (see Supplementary Materials). In other words, this set-up is exclusively sensitive to the octupolar degrees of freedom in the system, at least to the extent that thermal population of the higher energy excited CEF states can be neglected. 

The octupolar fluctuations are then probed by the AC elastocaloric effect \cite{MatthiasBase,MatthiasSusceptibility,Li2022-db}. We use a commercial Razorbill CS100 strain cell to deform the sample, and monitor the displacement of the jaws of the strain cell using a capacitor sensor, from which changes in the longitudinal strain in the [111] direction, which we denote as $\varepsilon_{111}\equiv\Delta L_{111}/L_{111}$, are extracted. The three shear strains, $\epsilon_{xy}$, $\epsilon_{yz}$ and $\epsilon_{zx}$, are equal to each other and proportional to $\varepsilon_{111}$ (see Supplementary Materials). In the measurement, a static (DC) offset strain is applied to the sample - this arises in part from differential thermal contraction of the material with respect to the materials of the strain cell (unknown in magnitude, meaning that the true `zero strain point' is unknown) and in part from controlled variation of the strain cell (meaning that we precisely know variation of the strain with respect to this unknown starting point). A small-amplitude AC strain is superimposed (typically of order $10^{-5}-10^{-4}$) at a sufficiently high frequency that heat flow in/out of the sample is suppressed and hence these AC sample deformations occur under a quasi-adiabatic condition. Changes in the strain-induced splitting of the CEF doublet then necessarily induce temperature oscillations at the  same frequency in order to keep the total entropy constant. The adiabatic elastocaloric temperature oscillation is thus given by \cite{MatthiasBase}
\begin{equation}
    \label{signalEquation}
    \left(\frac{\partial T}{\partial\epsilon}\right)_{S} = -\frac{T}{C_\epsilon}\left(\frac{\partial S}{\partial \epsilon}\right)_T
\end{equation}
which, as we will show, offers a window into the octupole susceptibility. 

In order to illustrate the expected elastocaloric response we consider a simplified Landau model, in which the free energy is given by 
\begin{equation}\label{free_energy_0}
    F = \frac{a}{2} O^2 - \lambda_1 H \epsilon O - \lambda_2 H^3 O
\end{equation}
where $O$ represents the total octupolar moment of the system, and $a=a_0(T-T^*)$ the $T$-dependent Landau free energy parameter with $T^*$ being the octupolar ordering temperature. $\lambda_1H\epsilon O$ ($\lambda_2H^3O$) describes the allowed bilinear coupling between the octupolar moments and $H_i\epsilon_{jk}$ ($H_iH_jH_k$). $H$ indicates the strength of the [111]-oriented magnetic field and $\epsilon$ the strength of shear strain forming $H_i\epsilon_{jk}$ with $H$ ($\epsilon$ is directly proportional to the measured/controlled $\varepsilon_{111}$, see Supplemental Material). This model is relevant for sufficiently high temperatures above any octupolar (or competing) phase transition, and for sufficiently small values of $H$ and $\epsilon$ (for instance such that the effect of coupling to potential quadrupolar fluctuations can be integrated out to yield such an `octupolar-only' effective theory). To illustrate the evolution of $O$ we combine the $\lambda_1$ and $\lambda_2$ terms in Eq. \ref{free_energy_0} for simplicity (absorbing a factor into the definition of $\epsilon$):
\begin{equation}\label{free_energy}
    F = \frac{a}{2} O^2 - \lambda H (\epsilon-H^2) O
\end{equation}
Inspection of Eqn. 3 reveals that for non-zero $H$ the octupolar moment will be zero not at zero applied strain, but at some finite strain wherein the two allowed couplings cancel. Minimizing Eq. \ref{free_energy} with respect to $O$ yields $O = \lambda H (\epsilon - H^2)/a$ and hence $F = -\lambda^2 H^2 (\epsilon-H^2)^2/2a$. From this, an octupolar susceptibility with respect to the composite strain-magnetic field can be defined:
\begin{equation}\label{chi_O}
    \chi_O \equiv\left. \frac{\partial O}{\partial(H\epsilon)}\right\rvert_{O = 0} = \frac{\lambda}{a}
\end{equation}
Although susceptibilities are usually defined in the limit of both vanishing conjugate field and vanishing order parameter, given the inequivalence of those two points ($O\rightarrow0$ and $H\epsilon\rightarrow0$) in the phase space in the present case, and in particular given the challenge of precisely identifying the true `zero strain' point, we suggest the limit of vanishing order parameter ($O\rightarrow0$) to be more physically relevant and more experimentally feasible, as we return to below.

$S$ can then be solved for, followed by $dS/d\epsilon$
\begin{equation}
    S = -\frac{\partial F}{\partial T} = \frac{d\chi_O}{dT}\frac{\lambda H^2 (\epsilon-H^2)^2}{2}
\end{equation}
\begin{equation}
    \frac{\partial S}{\partial \epsilon} = \lambda \frac{d\chi_O}{dT} H^2 (\epsilon-H^2)
    \label{entropyDerivativeEquation}
\end{equation}
Thus, for small perturbations from the zero-octupole point, and sufficiently far away from any phase transition, the elastocaloric coefficient $dT/d\epsilon$ is anticipated to be linearly dependent on $\epsilon$, with a slope that is quadratic in $H$. Determination of the coefficient of the strain- and field-dependence of the elastocaloric effect as a function of temperature then allows the extraction of the temperature-dependence of $d\chi_O/dT$ and thus of $\chi_O$. The mean-field expectation for $\chi_O$ based on Eqn. 2 is a Curie-Weiss functional form, and thus $d\chi_O/dT$ is anticipated to follow $(T-T^*)^{-2}$, as has been previously observed for nematic and quadrupolar-strain susceptibilities measured by a similar strategy \cite{chu2012divergent,Elliott,MatthiasSusceptibility}. 

As a starting point for examining the octupolar response in \ce{PrV2Al20}, the elastocaloric coefficient $dT/d\varepsilon_{111}$ is examined at various fields/temperatures (Fig. \ref{fig-2}). In Fig. \ref{fig-2}(a) we show $dT/d\varepsilon_{111}$ as a function of $\varepsilon_{111}$, measured with and without a magnetic field at $T=2.5$ K.  In the absence of a magnetic field, the signal is small and approximately linear w.r.t. strain, likely due to the small $\Gamma_5$ quadrupolar susceptibility originating from excited crystal field states mentioned above. With increasing $H$, the slope of $dT/d\varepsilon_{111}$ as a function of $\varepsilon_{111}$ is found to rapidly increase, as expected from the Landau model (Eqs. 2-6). Thus, the assumption of CEF-forbidden contributions being much weaker than the octupolar contribution is made for the remainder of the analysis. We also note that the zero intercept of $dT/d\varepsilon_{111}$ curves in Fig. \ref{fig-2}(a) indeed vary with $H$, representing a shift in the strain required for cancellation of $H\epsilon$ and $H^3$ couplings with each other as a function of field discussed above.

\begin{figure*}
\centering
\includegraphics[width =1.5 \columnwidth]{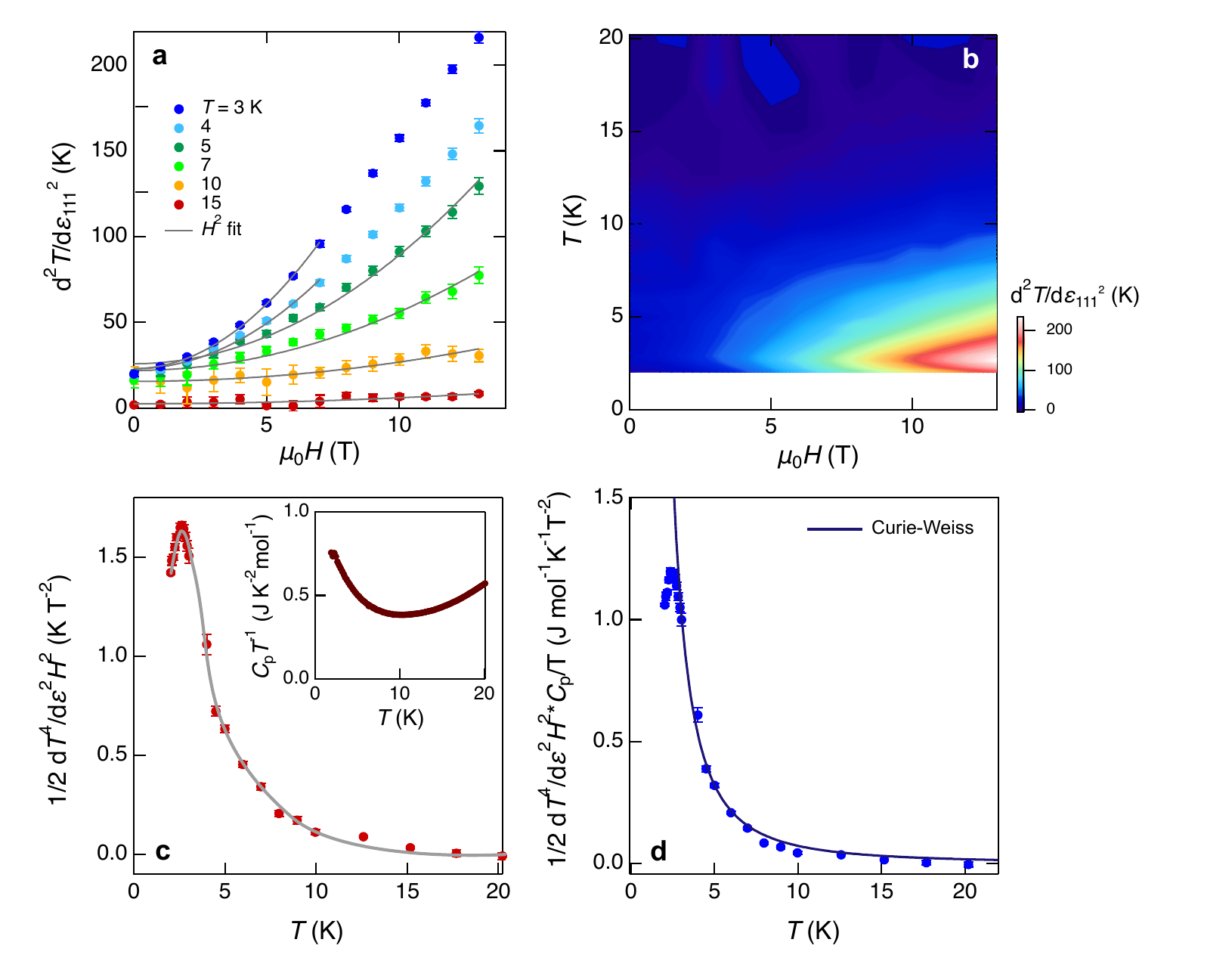}
\caption{\textbf{Extracting octupolar susceptibility from the elastocaloric coefficients} (a) Field-dependence of the strain-slope of $dT/d\varepsilon_{111}$ ($d^2T/d\varepsilon_{111}^2$) evaluated near zero $dT/d\varepsilon_{111}$ (\textit{i.e.} at approximate zero octupole moment; see text) for representative $T$ (solid symbols). The quadratic fit to $d^2T/d\varepsilon_{111}^2$ at each $T$ is shown as gray solid lines. (b) Color plot of $d^2T/d\varepsilon_{111}^2$ in the temperature-magnetic field plane. The corresponding color scale is shown on the bottom left. (c) Extracted $1/2d^4T/d\varepsilon^2 d H^2$ as a function of $T$; the gray curve is a guide to the eye. Inset shows $C_p/T$ of \ce{PrV2Al20}. (d) $1/2(d^4T/d\varepsilon^2 d H^2)(C_p/T)$ as a function of $T$.  The solid line indicates a Curie-Weiss fit of the form $(T-T^*)^{-2}$ (see text). }
\label{fig-3}
\end{figure*}

The bias-strain-dependence of $dT/d\varepsilon_{111}$ at fixed field ($\mu_0H= 10$ T) and for selected representative temperatures is shown in Fig. \ref{fig-2}(b), revealing a slope that rapidly increases upon cooling. It should be noted that the traces  in both Fig. \ref{fig-2}(a) and (b) become significantly more curved as their average slope increases at lower $T$ and higher $H$. This can be attributed to $\epsilon$-non-linearity when $O$ and the corresponding field-induced shear quadrupoles (see Fig. \ref{fig-1}(b), $\propto OH$) are far from zero. The simple Landau picture described above should nevertheless hold in the region where $O\rightarrow0$, which is reasonably approximated (but not perfectly matched, due to potential effects from non-symmetry-breaking strain) by the zero of $dT/d\varepsilon_{111}$. This motivates taking the strain-derivative of $dT/d\varepsilon_{111}$ in the limit $dT/d\varepsilon_{111}\rightarrow0$. To do this, $dT/d\varepsilon_{111}$ as a function of $\varepsilon_{111}$ was fit for each temperature with a 2$^{\text{nd}}$-order polynomial and the derivative evaluated at the polynomial's zero intercept to extract $d^2T/d\varepsilon_{111}^2$ in the limit of vanishing order parameter as well as vanishing higher-order contributions. At higher $T$ (above 10K) where $dT/d\varepsilon_{111}$ is linear with $\varepsilon_{111}$ (see Supplementary Materials) we simply use the slope of the linear fits to $dT/d\varepsilon_{111}$ to determine $d^2T/d\varepsilon_{111}^2$. 

The extracted $d^2T/d\varepsilon_{111}^2$ with $H$ is summarized in Fig. \ref{fig-3}(a) and its evolution over the entire $H-T$ plane in Fig. \ref{fig-3}(b); it is apparent that $d^2T/d\varepsilon_{111}^2$ increases monotonically with $H$ and the most prominent response can be found at low $T$ and high $H$. The data at each temperature were fit to $\alpha + \beta H^2$ where $\alpha$ and $\beta$ are constants (Fig. \ref{fig-3}(a)), with the small $\alpha$ coefficient deriving from the $\Gamma_5$ quadrupole moments described earlier (see Supplementary Materials for a discussion on the range of fitting used to extract $\alpha$ and $\beta$). The quadratic coefficient $\beta$ obtained by this procedure for each $T$ is summarized in Fig. \ref{fig-3}(c): an onset of measurable signal is observed from below approximately 15 K, and exhibits a strong increase with decreasing $T$, before rolling over below 2.5 K. The characteristic `fishtail' shape has been observed for similar low-temperature elastocaloric measurements of the quadrupolar material TmVO$_4$ \cite{zic2023giant} and is anticipated for an entropy landscape in which contributions to the total entropy arise from the splitting of the CEF groundstate (here induced by the $H_i\epsilon_{jk}$ effective field) as well as the largely strain-independent phonon contribution.

According to the thermodynamic argument presented earlier, in the weak-field limit the $H^2$ coefficient reflects the combined effects of the octupolar susceptibility and the heat capacity via
\begin{equation}
    \frac{d^2T}{d\epsilon^2} = -\frac{\partial^2 S}{\partial \epsilon^2}\frac{T}{C_\epsilon} = -\frac{T}{C_\epsilon}\frac{d\chi_O}{dT} H^2
\end{equation} 
Hence, we can extract $d\chi_O/dT$ from the $\beta$ coefficients  by multiplying by $C_\epsilon /T$. 

We use the zero field $C_p/T$ (Fig. \ref{fig-3}(c) inset) as a close approximation of $C_\epsilon$, since the $\Gamma_3$ doublet in principle remains un-split in this case, which should also hold true near the zero octupole point. In practice, subtle strain effects associated with the heat capacity measurements (\textit{i.e.} strain-induced splitting of the CEF multiplet that arise when the sample is held on a sample platform for heat capacity measurements) may potentially affect the lowest temperature data. The resulting estimate of $d\chi_O/dT$ as a function of $T$ is shown in the main panel of Fig. \ref{fig-3}(d). 

Fitting data away from the low-T saturation ($T>3$ K) to $(T-T^*)^{-2}$ expected for $d\chi_O/dT$ (black solid line) yields a remarkably good fit to this Curie-Weiss form, with $T^*=(0.5\pm0.15)$ K and a Curie constant $C=(6.5\pm0.7)$ J/mol/K/T$^2$ (dependence of the fitting results on the temperature range see Supplementary Materials). Similar $(T-T^*)^{-2}$ behavior has been previously observed for other entropy-based detection of susceptibility responses of lower rank multipoles using magnetocaloric \cite{oesterreicher1984magnetic} and elastocaloric effects \cite{MatthiasSusceptibility,rosenberg2023nematic,zic2023giant}. 

Deviations from the para-octupolar Curie behavior below approximately 3 K might reflect a variety of possible scenarios.  Subtle strain-dependence of the heat capacity mentioned above could affect the deduced temperature dependence of $\chi_O$ at low temperatures \cite{zic2023giant}. The effect might also reflect more complicated interactions with the competing order in the quadrupolar channel known to be present in \ce{PrV2Al20} \cite{Tsujimoto2014-cm}, or might even be related to the onset of a low-temperature multipolar Kondo effect, partially screening the growing octupole susceptibility \cite{Sakai2011-ks,patri2020critical,Santini2009-tn}. These possibilities, which, importantly, do not affect the key observation of a Curie-Weiss octupolar susceptibility at higher temperatures, can potentially be resolved by high-precision low-temperature heat capacity measurements performed as a function of strain.

The robust observation of a diverging octupole susceptibility up to temperatures approximately a factor of 40 greater than the critical temperature clearly reveals the presence of significant octupolar fluctuations in \ce{PrV2Al20}. Uncertainty in the Weiss temperature $T^*$ relative to the absolute value are quite large, but two standard deviations includes both observed critical temperatures, implying that octupolar order is potentially a viable competing state in \ce{PrV2Al20}, if not perhaps even realized at the lower temperature phase transition \cite{Tsujimoto2014-cm}. 

The composite nature of the effective field $H_i\epsilon_{jk}$ used here allows a simple protocol to isolate magnetic octupolar contributions to the total entropy, providing insight to the fluctuations of an otherwise elusive `hidden' order. More broadly, we anticipate the tools and methodology demonstrated in this study can be used to explore fluctuations (and be naturally extended to also probe and manipulate the corresponding ordered states) in the wider class of noncollinear antiferromagnets \cite{Suzuki2017-oz} and altermagnets \cite{vsmejkal2022emerging,bhowal2022magnetic}, along with other novel types of hidden order following similar symmetry-based arguments \cite{aeppli2020hidden}.\\

We acknowledge fruitful discussions with M. Ikeda, Y. B. Kim and A. Paramekanti. This work was supported by the Gordon and Betty Moore Foundation Emergent Phenomena in Quantum Systems Initiative through Grant GBMF9068. L.Y. acknowledges partial support from the National Science Foundation (DMR2232515) and the Marvin Chodorow Postdoctoral Fellowship at the Department of Applied Physics, Stanford University.

\bibliography{Bib.bib}

\begin{thebibliography}{35}%
\makeatletter
\providecommand \@ifxundefined [1]{%
 \@ifx{#1\undefined}
}%
\providecommand \@ifnum [1]{%
 \ifnum #1\expandafter \@firstoftwo
 \else \expandafter \@secondoftwo
 \fi
}%
\providecommand \@ifx [1]{%
 \ifx #1\expandafter \@firstoftwo
 \else \expandafter \@secondoftwo
 \fi
}%
\providecommand \natexlab [1]{#1}%
\providecommand \enquote  [1]{``#1''}%
\providecommand \bibnamefont  [1]{#1}%
\providecommand \bibfnamefont [1]{#1}%
\providecommand \citenamefont [1]{#1}%
\providecommand \href@noop [0]{\@secondoftwo}%
\providecommand \href [0]{\begingroup \@sanitize@url \@href}%
\providecommand \@href[1]{\@@startlink{#1}\@@href}%
\providecommand \@@href[1]{\endgroup#1\@@endlink}%
\providecommand \@sanitize@url [0]{\catcode `\\12\catcode `\$12\catcode `\&12\catcode `\#12\catcode `\^12\catcode `\_12\catcode `\%12\relax}%
\providecommand \@@startlink[1]{}%
\providecommand \@@endlink[0]{}%
\providecommand \url  [0]{\begingroup\@sanitize@url \@url }%
\providecommand \@url [1]{\endgroup\@href {#1}{\urlprefix }}%
\providecommand \urlprefix  [0]{URL }%
\providecommand \Eprint [0]{\href }%
\providecommand \doibase [0]{https://doi.org/}%
\providecommand \selectlanguage [0]{\@gobble}%
\providecommand \bibinfo  [0]{\@secondoftwo}%
\providecommand \bibfield  [0]{\@secondoftwo}%
\providecommand \translation [1]{[#1]}%
\providecommand \BibitemOpen [0]{}%
\providecommand \bibitemStop [0]{}%
\providecommand \bibitemNoStop [0]{.\EOS\space}%
\providecommand \EOS [0]{\spacefactor3000\relax}%
\providecommand \BibitemShut  [1]{\csname bibitem#1\endcsname}%
\let\auto@bib@innerbib\@empty
\bibitem [{\citenamefont {Santini}\ \emph {et~al.}(2009)\citenamefont {Santini}, \citenamefont {Carretta}, \citenamefont {Amoretti}, \citenamefont {Caciuffo}, \citenamefont {Magnani},\ and\ \citenamefont {Lander}}]{Santini2009-tn}%
  \BibitemOpen
  \bibfield  {author} {\bibinfo {author} {\bibfnamefont {P.}~\bibnamefont {Santini}}, \bibinfo {author} {\bibfnamefont {S.}~\bibnamefont {Carretta}}, \bibinfo {author} {\bibfnamefont {G.}~\bibnamefont {Amoretti}}, \bibinfo {author} {\bibfnamefont {R.}~\bibnamefont {Caciuffo}}, \bibinfo {author} {\bibfnamefont {N.}~\bibnamefont {Magnani}},\ and\ \bibinfo {author} {\bibfnamefont {G.~H.}\ \bibnamefont {Lander}},\ }\bibfield  {title} {\bibinfo {title} {Multipolar interactions inf-electron systems: The paradigm of actinide dioxides},\ }\href@noop {} {\bibfield  {journal} {\bibinfo  {journal} {Rev. Mod. Phys.}\ }\textbf {\bibinfo {volume} {81}},\ \bibinfo {pages} {807} (\bibinfo {year} {2009})}\BibitemShut {NoStop}%
\bibitem [{\citenamefont {Matsumura}(2017)}]{Matsumura2017-tm}%
  \BibitemOpen
  \bibfield  {author} {\bibinfo {author} {\bibfnamefont {T.}~\bibnamefont {Matsumura}},\ }\bibfield  {title} {\bibinfo {title} {Observation of multipole orderings in f-electron systems by resonant x-ray diffraction},\ }in\ \href@noop {} {\emph {\bibinfo {booktitle} {Resonant {X-Ray} Scattering in Correlated Systems}}},\ \bibinfo {editor} {edited by\ \bibinfo {editor} {\bibfnamefont {Y.}~\bibnamefont {Murakami}}\ and\ \bibinfo {editor} {\bibfnamefont {S.}~\bibnamefont {Ishihara}}}\ (\bibinfo  {publisher} {Springer Berlin Heidelberg},\ \bibinfo {address} {Berlin, Heidelberg},\ \bibinfo {year} {2017})\ pp.\ \bibinfo {pages} {85--117}\BibitemShut {NoStop}%
\bibitem [{\citenamefont {Sibille}\ \emph {et~al.}(2020)\citenamefont {Sibille}, \citenamefont {Gauthier}, \citenamefont {Lhotel}, \citenamefont {Por{\'e}e}, \citenamefont {Pomjakushin}, \citenamefont {Ewings}, \citenamefont {Perring}, \citenamefont {Ollivier}, \citenamefont {Wildes}, \citenamefont {Ritter}, \citenamefont {Hansen}, \citenamefont {Keen}, \citenamefont {Nilsen}, \citenamefont {Keller}, \citenamefont {Petit},\ and\ \citenamefont {Fennell}}]{Sibille2020-qd}%
  \BibitemOpen
  \bibfield  {author} {\bibinfo {author} {\bibfnamefont {R.}~\bibnamefont {Sibille}}, \bibinfo {author} {\bibfnamefont {N.}~\bibnamefont {Gauthier}}, \bibinfo {author} {\bibfnamefont {E.}~\bibnamefont {Lhotel}}, \bibinfo {author} {\bibfnamefont {V.}~\bibnamefont {Por{\'e}e}}, \bibinfo {author} {\bibfnamefont {V.}~\bibnamefont {Pomjakushin}}, \bibinfo {author} {\bibfnamefont {R.~A.}\ \bibnamefont {Ewings}}, \bibinfo {author} {\bibfnamefont {T.~G.}\ \bibnamefont {Perring}}, \bibinfo {author} {\bibfnamefont {J.}~\bibnamefont {Ollivier}}, \bibinfo {author} {\bibfnamefont {A.}~\bibnamefont {Wildes}}, \bibinfo {author} {\bibfnamefont {C.}~\bibnamefont {Ritter}}, \bibinfo {author} {\bibfnamefont {T.~C.}\ \bibnamefont {Hansen}}, \bibinfo {author} {\bibfnamefont {D.~A.}\ \bibnamefont {Keen}}, \bibinfo {author} {\bibfnamefont {G.~J.}\ \bibnamefont {Nilsen}}, \bibinfo {author} {\bibfnamefont {L.}~\bibnamefont {Keller}}, \bibinfo {author} {\bibfnamefont {S.}~\bibnamefont {Petit}},\ and\ \bibinfo {author} {\bibfnamefont
  {T.}~\bibnamefont {Fennell}},\ }\bibfield  {title} {\bibinfo {title} {A quantum liquid of magnetic octupoles on the pyrochlore lattice},\ }\href@noop {} {\bibfield  {journal} {\bibinfo  {journal} {Nat. Phys.}\ }\textbf {\bibinfo {volume} {16}},\ \bibinfo {pages} {546} (\bibinfo {year} {2020})}\BibitemShut {NoStop}%
\bibitem [{\citenamefont {Maharaj}\ \emph {et~al.}(2020)\citenamefont {Maharaj}, \citenamefont {Sala}, \citenamefont {Stone}, \citenamefont {Kermarrec}, \citenamefont {Ritter}, \citenamefont {Fauth}, \citenamefont {Marjerrison}, \citenamefont {Greedan}, \citenamefont {Paramekanti},\ and\ \citenamefont {Gaulin}}]{PhysRevLett.124.087206}%
  \BibitemOpen
  \bibfield  {author} {\bibinfo {author} {\bibfnamefont {D.~D.}\ \bibnamefont {Maharaj}}, \bibinfo {author} {\bibfnamefont {G.}~\bibnamefont {Sala}}, \bibinfo {author} {\bibfnamefont {M.~B.}\ \bibnamefont {Stone}}, \bibinfo {author} {\bibfnamefont {E.}~\bibnamefont {Kermarrec}}, \bibinfo {author} {\bibfnamefont {C.}~\bibnamefont {Ritter}}, \bibinfo {author} {\bibfnamefont {F.}~\bibnamefont {Fauth}}, \bibinfo {author} {\bibfnamefont {C.~A.}\ \bibnamefont {Marjerrison}}, \bibinfo {author} {\bibfnamefont {J.~E.}\ \bibnamefont {Greedan}}, \bibinfo {author} {\bibfnamefont {A.}~\bibnamefont {Paramekanti}},\ and\ \bibinfo {author} {\bibfnamefont {B.~D.}\ \bibnamefont {Gaulin}},\ }\bibfield  {title} {\bibinfo {title} {Octupolar versus n\'eel order in cubic $5{d}^{2}$ double perovskites},\ }\href@noop {} {\bibfield  {journal} {\bibinfo  {journal} {Phys. Rev. Lett.}\ }\textbf {\bibinfo {volume} {124}},\ \bibinfo {pages} {087206} (\bibinfo {year} {2020})}\BibitemShut {NoStop}%
\bibitem [{\citenamefont {Suzuki}\ \emph {et~al.}(2017)\citenamefont {Suzuki}, \citenamefont {Koretsune}, \citenamefont {Ochi},\ and\ \citenamefont {Arita}}]{Suzuki2017-oz}%
  \BibitemOpen
  \bibfield  {author} {\bibinfo {author} {\bibfnamefont {M.-T.}\ \bibnamefont {Suzuki}}, \bibinfo {author} {\bibfnamefont {T.}~\bibnamefont {Koretsune}}, \bibinfo {author} {\bibfnamefont {M.}~\bibnamefont {Ochi}},\ and\ \bibinfo {author} {\bibfnamefont {R.}~\bibnamefont {Arita}},\ }\bibfield  {title} {\bibinfo {title} {Cluster multipole theory for anomalous {Hall} effect in antiferromagnets},\ }\href@noop {} {\bibfield  {journal} {\bibinfo  {journal} {Phys. Rev. B}\ }\textbf {\bibinfo {volume} {95}},\ \bibinfo {pages} {094406} (\bibinfo {year} {2017})}\BibitemShut {NoStop}%
\bibitem [{\citenamefont {Wang}\ \emph {et~al.}(2017)\citenamefont {Wang}, \citenamefont {Weng}, \citenamefont {Fu},\ and\ \citenamefont {Dai}}]{Wang2017-om}%
  \BibitemOpen
  \bibfield  {author} {\bibinfo {author} {\bibfnamefont {Y.}~\bibnamefont {Wang}}, \bibinfo {author} {\bibfnamefont {H.}~\bibnamefont {Weng}}, \bibinfo {author} {\bibfnamefont {L.}~\bibnamefont {Fu}},\ and\ \bibinfo {author} {\bibfnamefont {X.}~\bibnamefont {Dai}},\ }\bibfield  {title} {\bibinfo {title} {Noncollinear magnetic structure and multipolar order in \ce{Eu2Ir2O7}},\ }\href@noop {} {\bibfield  {journal} {\bibinfo  {journal} {Phys. Rev. Lett.}\ }\textbf {\bibinfo {volume} {119}},\ \bibinfo {pages} {187203} (\bibinfo {year} {2017})}\BibitemShut {NoStop}%
\bibitem [{\citenamefont {Higo}\ \emph {et~al.}(2018)\citenamefont {Higo}, \citenamefont {Man}, \citenamefont {Gopman}, \citenamefont {Wu}, \citenamefont {Koretsune}, \citenamefont {van~'t Erve}, \citenamefont {Kabanov}, \citenamefont {Rees}, \citenamefont {Li}, \citenamefont {Suzuki}, \citenamefont {Patankar}, \citenamefont {Ikhlas}, \citenamefont {Chien}, \citenamefont {Arita}, \citenamefont {Shull}, \citenamefont {Orenstein},\ and\ \citenamefont {Nakatsuji}}]{Higo2018-oq}%
  \BibitemOpen
  \bibfield  {author} {\bibinfo {author} {\bibfnamefont {T.}~\bibnamefont {Higo}}, \bibinfo {author} {\bibfnamefont {H.}~\bibnamefont {Man}}, \bibinfo {author} {\bibfnamefont {D.~B.}\ \bibnamefont {Gopman}}, \bibinfo {author} {\bibfnamefont {L.}~\bibnamefont {Wu}}, \bibinfo {author} {\bibfnamefont {T.}~\bibnamefont {Koretsune}}, \bibinfo {author} {\bibfnamefont {O.~M.~J.}\ \bibnamefont {van~'t Erve}}, \bibinfo {author} {\bibfnamefont {Y.~P.}\ \bibnamefont {Kabanov}}, \bibinfo {author} {\bibfnamefont {D.}~\bibnamefont {Rees}}, \bibinfo {author} {\bibfnamefont {Y.}~\bibnamefont {Li}}, \bibinfo {author} {\bibfnamefont {M.-T.}\ \bibnamefont {Suzuki}}, \bibinfo {author} {\bibfnamefont {S.}~\bibnamefont {Patankar}}, \bibinfo {author} {\bibfnamefont {M.}~\bibnamefont {Ikhlas}}, \bibinfo {author} {\bibfnamefont {C.~L.}\ \bibnamefont {Chien}}, \bibinfo {author} {\bibfnamefont {R.}~\bibnamefont {Arita}}, \bibinfo {author} {\bibfnamefont {R.~D.}\ \bibnamefont {Shull}}, \bibinfo {author} {\bibfnamefont {J.}~\bibnamefont
  {Orenstein}},\ and\ \bibinfo {author} {\bibfnamefont {S.}~\bibnamefont {Nakatsuji}},\ }\bibfield  {title} {\bibinfo {title} {Large magneto-optical kerr effect and imaging of magnetic octupole domains in an antiferromagnetic metal},\ }\href@noop {} {\bibfield  {journal} {\bibinfo  {journal} {Nat. Photonics}\ }\textbf {\bibinfo {volume} {12}},\ \bibinfo {pages} {73} (\bibinfo {year} {2018})}\BibitemShut {NoStop}%
\bibitem [{\citenamefont {Nakatsuji}\ and\ \citenamefont {Arita}(2022)}]{nakatsuji2022topological}%
  \BibitemOpen
  \bibfield  {author} {\bibinfo {author} {\bibfnamefont {S.}~\bibnamefont {Nakatsuji}}\ and\ \bibinfo {author} {\bibfnamefont {R.}~\bibnamefont {Arita}},\ }\bibfield  {title} {\bibinfo {title} {Topological magnets: Functions based on berry phase and multipoles},\ }\href@noop {} {\bibfield  {journal} {\bibinfo  {journal} {Annual Review of Condensed Matter Physics}\ }\textbf {\bibinfo {volume} {13}},\ \bibinfo {pages} {119} (\bibinfo {year} {2022})}\BibitemShut {NoStop}%
\bibitem [{\citenamefont {{\v{S}}mejkal}\ \emph {et~al.}(2022)\citenamefont {{\v{S}}mejkal}, \citenamefont {Sinova},\ and\ \citenamefont {Jungwirth}}]{vsmejkal2022emerging}%
  \BibitemOpen
  \bibfield  {author} {\bibinfo {author} {\bibfnamefont {L.}~\bibnamefont {{\v{S}}mejkal}}, \bibinfo {author} {\bibfnamefont {J.}~\bibnamefont {Sinova}},\ and\ \bibinfo {author} {\bibfnamefont {T.}~\bibnamefont {Jungwirth}},\ }\bibfield  {title} {\bibinfo {title} {Emerging research landscape of altermagnetism},\ }\href@noop {} {\bibfield  {journal} {\bibinfo  {journal} {Physical Review X}\ }\textbf {\bibinfo {volume} {12}},\ \bibinfo {pages} {040501} (\bibinfo {year} {2022})}\BibitemShut {NoStop}%
\bibitem [{\citenamefont {Bhowal}\ and\ \citenamefont {Spaldin}(2022)}]{bhowal2022magnetic}%
  \BibitemOpen
  \bibfield  {author} {\bibinfo {author} {\bibfnamefont {S.}~\bibnamefont {Bhowal}}\ and\ \bibinfo {author} {\bibfnamefont {N.~A.}\ \bibnamefont {Spaldin}},\ }\bibfield  {title} {\bibinfo {title} {Magnetic octupoles as the order parameter for unconventional antiferromagnetism},\ }\href@noop {} {\bibfield  {journal} {\bibinfo  {journal} {arXiv preprint arXiv:2212.03756}\ } (\bibinfo {year} {2022})}\BibitemShut {NoStop}%
\bibitem [{\citenamefont {Onimaru}\ \emph {et~al.}(2010)\citenamefont {Onimaru}, \citenamefont {T.~Matsumoto}, \citenamefont {F.~Inoue}, \citenamefont {Umeo}, \citenamefont {Saiga}, \citenamefont {Matsushita}, \citenamefont {Tamura}, \citenamefont {Nishimoto}, \citenamefont {Ishii}, \citenamefont {Suzuki},\ and\ \citenamefont {Takabatake}}]{Onimaru2010-bm}%
  \BibitemOpen
  \bibfield  {author} {\bibinfo {author} {\bibfnamefont {T.}~\bibnamefont {Onimaru}}, \bibinfo {author} {\bibfnamefont {K.}~\bibnamefont {T.~Matsumoto}}, \bibinfo {author} {\bibfnamefont {Y.}~\bibnamefont {F.~Inoue}}, \bibinfo {author} {\bibfnamefont {K.}~\bibnamefont {Umeo}}, \bibinfo {author} {\bibfnamefont {Y.}~\bibnamefont {Saiga}}, \bibinfo {author} {\bibfnamefont {Y.}~\bibnamefont {Matsushita}}, \bibinfo {author} {\bibfnamefont {R.}~\bibnamefont {Tamura}}, \bibinfo {author} {\bibfnamefont {K.}~\bibnamefont {Nishimoto}}, \bibinfo {author} {\bibfnamefont {I.}~\bibnamefont {Ishii}}, \bibinfo {author} {\bibfnamefont {T.}~\bibnamefont {Suzuki}},\ and\ \bibinfo {author} {\bibfnamefont {T.}~\bibnamefont {Takabatake}},\ }\bibfield  {title} {\bibinfo {title} {Superconductivity and structural phase transitions in caged compounds \ce{RT2Zn20} (\ce{R} = \ce{La, Pr}, {T} = \ce{Ru, Ir})},\ }\href@noop {} {\bibfield  {journal} {\bibinfo  {journal} {J. Phys. Soc. Jpn.}\ }\textbf {\bibinfo {volume} {79}},\ \bibinfo
  {pages} {033704} (\bibinfo {year} {2010})}\BibitemShut {NoStop}%
\bibitem [{\citenamefont {Sakai}\ and\ \citenamefont {Nakatsuji}(2011)}]{Sakai2011-ks}%
  \BibitemOpen
  \bibfield  {author} {\bibinfo {author} {\bibfnamefont {A.}~\bibnamefont {Sakai}}\ and\ \bibinfo {author} {\bibfnamefont {S.}~\bibnamefont {Nakatsuji}},\ }\bibfield  {title} {\bibinfo {title} {Kondo effects and multipolar order in the cubic \ce{PrTr2Al20} ({\ce{tr}=\ce{ti}}, \ce{V})},\ }\href@noop {} {\bibfield  {journal} {\bibinfo  {journal} {J. Phys. Soc. Jpn.}\ }\textbf {\bibinfo {volume} {80}},\ \bibinfo {pages} {063701} (\bibinfo {year} {2011})}\BibitemShut {NoStop}%
\bibitem [{\citenamefont {Sakai}\ \emph {et~al.}(2012)\citenamefont {Sakai}, \citenamefont {Kuga},\ and\ \citenamefont {Nakatsuji}}]{Sakai2012-in}%
  \BibitemOpen
  \bibfield  {author} {\bibinfo {author} {\bibfnamefont {A.}~\bibnamefont {Sakai}}, \bibinfo {author} {\bibfnamefont {K.}~\bibnamefont {Kuga}},\ and\ \bibinfo {author} {\bibfnamefont {S.}~\bibnamefont {Nakatsuji}},\ }\bibfield  {title} {\bibinfo {title} {Superconductivity in the ferroquadrupolar state in the quadrupolar kondo lattice \ce{PrTi2Al20}},\ }\href@noop {} {\bibfield  {journal} {\bibinfo  {journal} {J. Phys. Soc. Jpn.}\ }\textbf {\bibinfo {volume} {81}},\ \bibinfo {pages} {083702} (\bibinfo {year} {2012})}\BibitemShut {NoStop}%
\bibitem [{\citenamefont {Tsujimoto}\ \emph {et~al.}(2014{\natexlab{a}})\citenamefont {Tsujimoto}, \citenamefont {Matsumoto}, \citenamefont {Tomita}, \citenamefont {Sakai},\ and\ \citenamefont {Nakatsuji}}]{Tsujimoto2014-cm}%
  \BibitemOpen
  \bibfield  {author} {\bibinfo {author} {\bibfnamefont {M.}~\bibnamefont {Tsujimoto}}, \bibinfo {author} {\bibfnamefont {Y.}~\bibnamefont {Matsumoto}}, \bibinfo {author} {\bibfnamefont {T.}~\bibnamefont {Tomita}}, \bibinfo {author} {\bibfnamefont {A.}~\bibnamefont {Sakai}},\ and\ \bibinfo {author} {\bibfnamefont {S.}~\bibnamefont {Nakatsuji}},\ }\bibfield  {title} {\bibinfo {title} {{Heavy-fermion} superconductivity in the quadrupole ordered state of \ce{PrV2Al20}},\ }\href@noop {} {\bibfield  {journal} {\bibinfo  {journal} {Phys. Rev. Lett.}\ }\textbf {\bibinfo {volume} {113}},\ \bibinfo {pages} {267001} (\bibinfo {year} {2014}{\natexlab{a}})}\BibitemShut {NoStop}%
\bibitem [{\citenamefont {Onimaru}\ \emph {et~al.}(2011)\citenamefont {Onimaru}, \citenamefont {Matsumoto}, \citenamefont {Inoue}, \citenamefont {Umeo}, \citenamefont {Sakakibara}, \citenamefont {Karaki}, \citenamefont {Kubota},\ and\ \citenamefont {Takabatake}}]{Onimaru2011-or}%
  \BibitemOpen
  \bibfield  {author} {\bibinfo {author} {\bibfnamefont {T.}~\bibnamefont {Onimaru}}, \bibinfo {author} {\bibfnamefont {K.~T.}\ \bibnamefont {Matsumoto}}, \bibinfo {author} {\bibfnamefont {Y.~F.}\ \bibnamefont {Inoue}}, \bibinfo {author} {\bibfnamefont {K.}~\bibnamefont {Umeo}}, \bibinfo {author} {\bibfnamefont {T.}~\bibnamefont {Sakakibara}}, \bibinfo {author} {\bibfnamefont {Y.}~\bibnamefont {Karaki}}, \bibinfo {author} {\bibfnamefont {M.}~\bibnamefont {Kubota}},\ and\ \bibinfo {author} {\bibfnamefont {T.}~\bibnamefont {Takabatake}},\ }\bibfield  {title} {\bibinfo {title} {Antiferroquadrupolar ordering in a {Pr-Based} superconductor \ce{PrIr2Zn20}},\ }\href@noop {} {\bibfield  {journal} {\bibinfo  {journal} {Phys. Rev. Lett.}\ }\textbf {\bibinfo {volume} {106}},\ \bibinfo {pages} {177001} (\bibinfo {year} {2011})}\BibitemShut {NoStop}%
\bibitem [{\citenamefont {Tsujimoto}\ \emph {et~al.}(2014{\natexlab{b}})\citenamefont {Tsujimoto}, \citenamefont {Matsumoto},\ and\ \citenamefont {Nakatsuji}}]{VHeatCapacity}%
  \BibitemOpen
  \bibfield  {author} {\bibinfo {author} {\bibfnamefont {M.}~\bibnamefont {Tsujimoto}}, \bibinfo {author} {\bibfnamefont {Y.}~\bibnamefont {Matsumoto}},\ and\ \bibinfo {author} {\bibfnamefont {S.}~\bibnamefont {Nakatsuji}},\ }\bibfield  {title} {\bibinfo {title} {Anomalous specific heat behaviour in the quadrupolar kondo system \ce{PrV2Al20}},\ }\href@noop {} {\bibfield  {journal} {\bibinfo  {journal} {J. Phys.: Conf. Ser.}\ }\textbf {\bibinfo {volume} {592}},\ \bibinfo {pages} {012023} (\bibinfo {year} {2014}{\natexlab{b}})}\BibitemShut {NoStop}%
\bibitem [{\citenamefont {Onimaru}\ and\ \citenamefont {Kusunose}(2016)}]{Onimaru2016-jz}%
  \BibitemOpen
  \bibfield  {author} {\bibinfo {author} {\bibfnamefont {T.}~\bibnamefont {Onimaru}}\ and\ \bibinfo {author} {\bibfnamefont {H.}~\bibnamefont {Kusunose}},\ }\bibfield  {title} {\bibinfo {title} {Exotic quadrupolar phenomena in {Non-Kramers} doublet systems --- the cases of \ce{PrT2Zn20} (t = \ce{Ir, Rh}) and \ce{PrT2Al20} (t = \ce{V,Ti}) ---},\ }\href@noop {} {\bibfield  {journal} {\bibinfo  {journal} {J. Phys. Soc. Jpn.}\ }\textbf {\bibinfo {volume} {85}},\ \bibinfo {pages} {082002} (\bibinfo {year} {2016})}\BibitemShut {NoStop}%
\bibitem [{\citenamefont {Lea}\ \emph {et~al.}(1962)\citenamefont {Lea}, \citenamefont {Leask},\ and\ \citenamefont {Wolf}}]{Lea1962-fp}%
  \BibitemOpen
  \bibfield  {author} {\bibinfo {author} {\bibfnamefont {K.~R.}\ \bibnamefont {Lea}}, \bibinfo {author} {\bibfnamefont {M.~J.~M.}\ \bibnamefont {Leask}},\ and\ \bibinfo {author} {\bibfnamefont {W.~P.}\ \bibnamefont {Wolf}},\ }\bibfield  {title} {\bibinfo {title} {The raising of angular momentum degeneracy of f-electron terms by cubic crystal fields},\ }\href@noop {} {\bibfield  {journal} {\bibinfo  {journal} {J. Phys. Chem. Solids}\ }\textbf {\bibinfo {volume} {23}},\ \bibinfo {pages} {1381} (\bibinfo {year} {1962})}\BibitemShut {NoStop}%
\bibitem [{\citenamefont {Freyer}\ \emph {et~al.}(2018)\citenamefont {Freyer}, \citenamefont {Attig}, \citenamefont {Lee}, \citenamefont {Paramekanti}, \citenamefont {Trebst},\ and\ \citenamefont {Kim}}]{Freyer2018-nf}%
  \BibitemOpen
  \bibfield  {author} {\bibinfo {author} {\bibfnamefont {F.}~\bibnamefont {Freyer}}, \bibinfo {author} {\bibfnamefont {J.}~\bibnamefont {Attig}}, \bibinfo {author} {\bibfnamefont {S.}~\bibnamefont {Lee}}, \bibinfo {author} {\bibfnamefont {A.}~\bibnamefont {Paramekanti}}, \bibinfo {author} {\bibfnamefont {S.}~\bibnamefont {Trebst}},\ and\ \bibinfo {author} {\bibfnamefont {Y.~B.}\ \bibnamefont {Kim}},\ }\bibfield  {title} {\bibinfo {title} {Two-stage multipolar ordering in \ce{PrT2Al20} kondo materials},\ }\href@noop {} {\bibfield  {journal} {\bibinfo  {journal} {Phys. Rev. B}\ }\textbf {\bibinfo {volume} {97}} (\bibinfo {year} {2018})}\BibitemShut {NoStop}%
\bibitem [{\citenamefont {Lee}\ \emph {et~al.}(2018)\citenamefont {Lee}, \citenamefont {Trebst}, \citenamefont {Kim},\ and\ \citenamefont {Paramekanti}}]{Lee2018-cn}%
  \BibitemOpen
  \bibfield  {author} {\bibinfo {author} {\bibfnamefont {S.}~\bibnamefont {Lee}}, \bibinfo {author} {\bibfnamefont {S.}~\bibnamefont {Trebst}}, \bibinfo {author} {\bibfnamefont {Y.~B.}\ \bibnamefont {Kim}},\ and\ \bibinfo {author} {\bibfnamefont {A.}~\bibnamefont {Paramekanti}},\ }\bibfield  {title} {\bibinfo {title} {Landau theory of multipolar orders in \ce{Pr(Y)2X20} kondo materials (\ce{Ti},\ce{V},\ce{Rh},\ce{Ir};\ce{Al},\ce{Zn})},\ }\href@noop {} {\bibfield  {journal} {\bibinfo  {journal} {Phys. Rev. B}\ }\textbf {\bibinfo {volume} {98}},\ \bibinfo {pages} {134447} (\bibinfo {year} {2018})}\BibitemShut {NoStop}%
\bibitem [{\citenamefont {Patri}\ \emph {et~al.}(2019)\citenamefont {Patri}, \citenamefont {Sakai}, \citenamefont {Lee}, \citenamefont {Paramekanti}, \citenamefont {Nakatsuji},\ and\ \citenamefont {Kim}}]{YongBaek}%
  \BibitemOpen
  \bibfield  {author} {\bibinfo {author} {\bibfnamefont {A.~S.}\ \bibnamefont {Patri}}, \bibinfo {author} {\bibfnamefont {A.}~\bibnamefont {Sakai}}, \bibinfo {author} {\bibfnamefont {S.}~\bibnamefont {Lee}}, \bibinfo {author} {\bibfnamefont {A.}~\bibnamefont {Paramekanti}}, \bibinfo {author} {\bibfnamefont {S.}~\bibnamefont {Nakatsuji}},\ and\ \bibinfo {author} {\bibfnamefont {Y.~B.}\ \bibnamefont {Kim}},\ }\bibfield  {title} {\bibinfo {title} {Unveiling hidden multipolar orders with magnetostriction},\ }\href@noop {} {\bibfield  {journal} {\bibinfo  {journal} {Nat. Commun.}\ }\textbf {\bibinfo {volume} {10}},\ \bibinfo {pages} {4092} (\bibinfo {year} {2019})}\BibitemShut {NoStop}%
\bibitem [{\citenamefont {Sorensen}\ and\ \citenamefont {Fisher}(2021)}]{SelfCiteGarbage}%
  \BibitemOpen
  \bibfield  {author} {\bibinfo {author} {\bibfnamefont {M.~E.}\ \bibnamefont {Sorensen}}\ and\ \bibinfo {author} {\bibfnamefont {I.~R.}\ \bibnamefont {Fisher}},\ }\bibfield  {title} {\bibinfo {title} {Proposal for methods to measure the octupole susceptibility in certain cubic \ce{Pr} compounds},\ }\href@noop {} {\bibfield  {journal} {\bibinfo  {journal} {Phys. Rev. B}\ }\textbf {\bibinfo {volume} {103}},\ \bibinfo {pages} {155106} (\bibinfo {year} {2021})}\BibitemShut {NoStop}%
\bibitem [{\citenamefont {L{\"u}thi}(2007)}]{luthi2007physical}%
  \BibitemOpen
  \bibfield  {author} {\bibinfo {author} {\bibfnamefont {B.}~\bibnamefont {L{\"u}thi}},\ }\href@noop {} {\emph {\bibinfo {title} {Physical acoustics in the solid state}}},\ Vol.\ \bibinfo {volume} {148}\ (\bibinfo  {publisher} {Springer Science \& Business Media},\ \bibinfo {year} {2007})\BibitemShut {NoStop}%
\bibitem [{\citenamefont {Ikeda}\ \emph {et~al.}(2019)\citenamefont {Ikeda}, \citenamefont {Straquadine}, \citenamefont {Hristov}, \citenamefont {Worasaran}, \citenamefont {Palmstrom}, \citenamefont {Sorensen}, \citenamefont {Walmsley},\ and\ \citenamefont {Fisher}}]{MatthiasBase}%
  \BibitemOpen
  \bibfield  {author} {\bibinfo {author} {\bibfnamefont {M.}~\bibnamefont {Ikeda}}, \bibinfo {author} {\bibfnamefont {J.}~\bibnamefont {Straquadine}}, \bibinfo {author} {\bibfnamefont {A.}~\bibnamefont {Hristov}}, \bibinfo {author} {\bibfnamefont {T.}~\bibnamefont {Worasaran}}, \bibinfo {author} {\bibfnamefont {J.}~\bibnamefont {Palmstrom}}, \bibinfo {author} {\bibfnamefont {M.}~\bibnamefont {Sorensen}}, \bibinfo {author} {\bibfnamefont {P.}~\bibnamefont {Walmsley}},\ and\ \bibinfo {author} {\bibfnamefont {I.}~\bibnamefont {Fisher}},\ }\bibfield  {title} {\bibinfo {title} {Ac elastocaloric effect as a probe for thermodynamic signatures of continuous phase transitions},\ }\href@noop {} {\bibfield  {journal} {\bibinfo  {journal} {Rev. Sci. Instrum.}\ }\textbf {\bibinfo {volume} {90}},\ \bibinfo {pages} {083902} (\bibinfo {year} {2019})}\BibitemShut {NoStop}%
\bibitem [{\citenamefont {Ikeda}\ \emph {et~al.}(2021)\citenamefont {Ikeda}, \citenamefont {Worasaran}, \citenamefont {Rosenberg}, \citenamefont {Palmstrom}, \citenamefont {Kivelson},\ and\ \citenamefont {Fisher}}]{MatthiasSusceptibility}%
  \BibitemOpen
  \bibfield  {author} {\bibinfo {author} {\bibfnamefont {M.}~\bibnamefont {Ikeda}}, \bibinfo {author} {\bibfnamefont {T.}~\bibnamefont {Worasaran}}, \bibinfo {author} {\bibfnamefont {E.}~\bibnamefont {Rosenberg}}, \bibinfo {author} {\bibfnamefont {J.}~\bibnamefont {Palmstrom}}, \bibinfo {author} {\bibfnamefont {S.}~\bibnamefont {Kivelson}},\ and\ \bibinfo {author} {\bibfnamefont {I.}~\bibnamefont {Fisher}},\ }\bibfield  {title} {\bibinfo {title} {Elastocaloric signature of nematic fluctuations},\ }\href@noop {} {\bibfield  {journal} {\bibinfo  {journal} {Proc. Natl. Acad. Sci. USA}\ }\textbf {\bibinfo {volume} {118(37)}},\ \bibinfo {pages} {e2105911118} (\bibinfo {year} {2021})}\BibitemShut {NoStop}%
\bibitem [{\citenamefont {Li}\ \emph {et~al.}(2022)\citenamefont {Li}, \citenamefont {Garst}, \citenamefont {Schmalian}, \citenamefont {Ghosh}, \citenamefont {Kikugawa}, \citenamefont {Sokolov}, \citenamefont {Hicks}, \citenamefont {Jerzembeck}, \citenamefont {Ikeda}, \citenamefont {Hu}, \citenamefont {Ramshaw}, \citenamefont {Rost}, \citenamefont {Nicklas},\ and\ \citenamefont {Mackenzie}}]{Li2022-db}%
  \BibitemOpen
  \bibfield  {author} {\bibinfo {author} {\bibfnamefont {Y.-S.}\ \bibnamefont {Li}}, \bibinfo {author} {\bibfnamefont {M.}~\bibnamefont {Garst}}, \bibinfo {author} {\bibfnamefont {J.}~\bibnamefont {Schmalian}}, \bibinfo {author} {\bibfnamefont {S.}~\bibnamefont {Ghosh}}, \bibinfo {author} {\bibfnamefont {N.}~\bibnamefont {Kikugawa}}, \bibinfo {author} {\bibfnamefont {D.~A.}\ \bibnamefont {Sokolov}}, \bibinfo {author} {\bibfnamefont {C.~W.}\ \bibnamefont {Hicks}}, \bibinfo {author} {\bibfnamefont {F.}~\bibnamefont {Jerzembeck}}, \bibinfo {author} {\bibfnamefont {M.~S.}\ \bibnamefont {Ikeda}}, \bibinfo {author} {\bibfnamefont {Z.}~\bibnamefont {Hu}}, \bibinfo {author} {\bibfnamefont {B.~J.}\ \bibnamefont {Ramshaw}}, \bibinfo {author} {\bibfnamefont {A.~W.}\ \bibnamefont {Rost}}, \bibinfo {author} {\bibfnamefont {M.}~\bibnamefont {Nicklas}},\ and\ \bibinfo {author} {\bibfnamefont {A.~P.}\ \bibnamefont {Mackenzie}},\ }\bibfield  {title} {\bibinfo {title} {Elastocaloric determination of the phase diagram of
  \ce{Sr2RuO4}},\ }\href@noop {} {\bibfield  {journal} {\bibinfo  {journal} {Nature}\ }\textbf {\bibinfo {volume} {607}},\ \bibinfo {pages} {276} (\bibinfo {year} {2022})}\BibitemShut {NoStop}%
\bibitem [{\citenamefont {Chu}\ \emph {et~al.}(2012)\citenamefont {Chu}, \citenamefont {Kuo}, \citenamefont {Analytis},\ and\ \citenamefont {Fisher}}]{chu2012divergent}%
  \BibitemOpen
  \bibfield  {author} {\bibinfo {author} {\bibfnamefont {J.-H.}\ \bibnamefont {Chu}}, \bibinfo {author} {\bibfnamefont {H.-H.}\ \bibnamefont {Kuo}}, \bibinfo {author} {\bibfnamefont {J.~G.}\ \bibnamefont {Analytis}},\ and\ \bibinfo {author} {\bibfnamefont {I.~R.}\ \bibnamefont {Fisher}},\ }\bibfield  {title} {\bibinfo {title} {Divergent nematic susceptibility in an iron arsenide superconductor},\ }\href@noop {} {\bibfield  {journal} {\bibinfo  {journal} {Science}\ }\textbf {\bibinfo {volume} {337}},\ \bibinfo {pages} {710} (\bibinfo {year} {2012})}\BibitemShut {NoStop}%
\bibitem [{\citenamefont {Rosenberg}\ \emph {et~al.}(2019)\citenamefont {Rosenberg}, \citenamefont {Chu}, \citenamefont {Ruff}, \citenamefont {Hristov},\ and\ \citenamefont {Fisher}}]{Elliott}%
  \BibitemOpen
  \bibfield  {author} {\bibinfo {author} {\bibfnamefont {E.~W.}\ \bibnamefont {Rosenberg}}, \bibinfo {author} {\bibfnamefont {J.-H.}\ \bibnamefont {Chu}}, \bibinfo {author} {\bibfnamefont {J.~P.}\ \bibnamefont {Ruff}}, \bibinfo {author} {\bibfnamefont {A.~T.}\ \bibnamefont {Hristov}},\ and\ \bibinfo {author} {\bibfnamefont {I.~R.}\ \bibnamefont {Fisher}},\ }\bibfield  {title} {\bibinfo {title} {Divergence of the quadrupole-strain susceptibility of the electronic nematic system \ce{YbRu2Ge2}},\ }\href@noop {} {\bibfield  {journal} {\bibinfo  {journal} {PNAS}\ }\textbf {\bibinfo {volume} {116 (15)}},\ \bibinfo {pages} {7232} (\bibinfo {year} {2019})}\BibitemShut {NoStop}%
\bibitem [{\citenamefont {Zic}\ \emph {et~al.}(2023)\citenamefont {Zic}, \citenamefont {Ikeda}, \citenamefont {Massat}, \citenamefont {Hollister}, \citenamefont {Ye}, \citenamefont {Rosenberg}, \citenamefont {Straquadine}, \citenamefont {Ramshaw},\ and\ \citenamefont {Fisher}}]{zic2023giant}%
  \BibitemOpen
  \bibfield  {author} {\bibinfo {author} {\bibfnamefont {M.~P.}\ \bibnamefont {Zic}}, \bibinfo {author} {\bibfnamefont {M.~S.}\ \bibnamefont {Ikeda}}, \bibinfo {author} {\bibfnamefont {P.}~\bibnamefont {Massat}}, \bibinfo {author} {\bibfnamefont {P.~M.}\ \bibnamefont {Hollister}}, \bibinfo {author} {\bibfnamefont {L.}~\bibnamefont {Ye}}, \bibinfo {author} {\bibfnamefont {E.~W.}\ \bibnamefont {Rosenberg}}, \bibinfo {author} {\bibfnamefont {J.~A.~W.}\ \bibnamefont {Straquadine}}, \bibinfo {author} {\bibfnamefont {B.~J.}\ \bibnamefont {Ramshaw}},\ and\ \bibinfo {author} {\bibfnamefont {I.~R.}\ \bibnamefont {Fisher}},\ }\href@noop {} {\bibinfo {title} {Giant elastocaloric effect at low temperatures in \ce{TmVO4} and implications for cryogenic cooling}} (\bibinfo {year} {2023})\BibitemShut {NoStop}%
\bibitem [{\citenamefont {Oesterreicher}\ and\ \citenamefont {Parker}(1984)}]{oesterreicher1984magnetic}%
  \BibitemOpen
  \bibfield  {author} {\bibinfo {author} {\bibfnamefont {H.}~\bibnamefont {Oesterreicher}}\ and\ \bibinfo {author} {\bibfnamefont {F.}~\bibnamefont {Parker}},\ }\bibfield  {title} {\bibinfo {title} {Magnetic cooling near curie temperatures above 300 {K}},\ }\href@noop {} {\bibfield  {journal} {\bibinfo  {journal} {Journal of applied physics}\ }\textbf {\bibinfo {volume} {55}},\ \bibinfo {pages} {4334} (\bibinfo {year} {1984})}\BibitemShut {NoStop}%
\bibitem [{\citenamefont {Rosenberg}\ \emph {et~al.}(2023)\citenamefont {Rosenberg}, \citenamefont {Ikeda},\ and\ \citenamefont {Fisher}}]{rosenberg2023nematic}%
  \BibitemOpen
  \bibfield  {author} {\bibinfo {author} {\bibfnamefont {E.~W.}\ \bibnamefont {Rosenberg}}, \bibinfo {author} {\bibfnamefont {M.}~\bibnamefont {Ikeda}},\ and\ \bibinfo {author} {\bibfnamefont {I.~R.}\ \bibnamefont {Fisher}},\ }\bibfield  {title} {\bibinfo {title} {The nematic susceptibility of the ferroquadrupolar metal \ce{TmAg2} measured via the elastocaloric effect},\ }\href@noop {} {\bibfield  {journal} {\bibinfo  {journal} {arXiv preprint arXiv:2308.05312}\ } (\bibinfo {year} {2023})}\BibitemShut {NoStop}%
\bibitem [{\citenamefont {Patri}\ and\ \citenamefont {Kim}(2020)}]{patri2020critical}%
  \BibitemOpen
  \bibfield  {author} {\bibinfo {author} {\bibfnamefont {A.~S.}\ \bibnamefont {Patri}}\ and\ \bibinfo {author} {\bibfnamefont {Y.~B.}\ \bibnamefont {Kim}},\ }\bibfield  {title} {\bibinfo {title} {Critical theory of non-fermi liquid fixed point in multipolar kondo problem},\ }\href@noop {} {\bibfield  {journal} {\bibinfo  {journal} {Physical Review X}\ }\textbf {\bibinfo {volume} {10}},\ \bibinfo {pages} {041021} (\bibinfo {year} {2020})}\BibitemShut {NoStop}%
\bibitem [{\citenamefont {Aeppli}\ \emph {et~al.}(2020)\citenamefont {Aeppli}, \citenamefont {Balatsky}, \citenamefont {R{\o}nnow},\ and\ \citenamefont {Spaldin}}]{aeppli2020hidden}%
  \BibitemOpen
  \bibfield  {author} {\bibinfo {author} {\bibfnamefont {G.}~\bibnamefont {Aeppli}}, \bibinfo {author} {\bibfnamefont {A.~V.}\ \bibnamefont {Balatsky}}, \bibinfo {author} {\bibfnamefont {H.~M.}\ \bibnamefont {R{\o}nnow}},\ and\ \bibinfo {author} {\bibfnamefont {N.~A.}\ \bibnamefont {Spaldin}},\ }\bibfield  {title} {\bibinfo {title} {Hidden, entangled and resonating order},\ }\href@noop {} {\bibfield  {journal} {\bibinfo  {journal} {Nat. Rev. Mater.}\ }\textbf {\bibinfo {volume} {5}},\ \bibinfo {pages} {477} (\bibinfo {year} {2020})}\BibitemShut {NoStop}%
\bibitem [{\citenamefont {Magata}\ \emph {et~al.}(2016)\citenamefont {Magata}, \citenamefont {Matsumoto}, \citenamefont {Tsujimoto}, \citenamefont {Tomita}, \citenamefont {Kiichler}, \citenamefont {Sakai},\ and\ \citenamefont {Nakatsuji}}]{magata2016low}%
  \BibitemOpen
  \bibfield  {author} {\bibinfo {author} {\bibfnamefont {A.}~\bibnamefont {Magata}}, \bibinfo {author} {\bibfnamefont {Y.}~\bibnamefont {Matsumoto}}, \bibinfo {author} {\bibfnamefont {M.}~\bibnamefont {Tsujimoto}}, \bibinfo {author} {\bibfnamefont {T.}~\bibnamefont {Tomita}}, \bibinfo {author} {\bibfnamefont {R.}~\bibnamefont {Kiichler}}, \bibinfo {author} {\bibfnamefont {A.}~\bibnamefont {Sakai}},\ and\ \bibinfo {author} {\bibfnamefont {S.}~\bibnamefont {Nakatsuji}},\ }\bibfield  {title} {\bibinfo {title} {Low-temperature thermal expansion measurements in \ce{PrV2Al20}},\ }in\ \href@noop {} {\emph {\bibinfo {booktitle} {Journal of Physics: Conference Series}}},\ Vol.\ \bibinfo {volume} {683}\ (\bibinfo {organization} {IOP Publishing},\ \bibinfo {year} {2016})\ p.\ \bibinfo {pages} {012014}\BibitemShut {NoStop}%
\bibitem [{\citenamefont {Ye}\ \emph {et~al.}(2023)\citenamefont {Ye}, \citenamefont {Sun}, \citenamefont {Sunko}, \citenamefont {Rodriguez-Nieva}, \citenamefont {Ikeda}, \citenamefont {Worasaran}, \citenamefont {Sorensen}, \citenamefont {Bachmann}, \citenamefont {Orenstein},\ and\ \citenamefont {Fisher}}]{ye2023elastocaloric}%
  \BibitemOpen
  \bibfield  {author} {\bibinfo {author} {\bibfnamefont {L.}~\bibnamefont {Ye}}, \bibinfo {author} {\bibfnamefont {Y.}~\bibnamefont {Sun}}, \bibinfo {author} {\bibfnamefont {V.}~\bibnamefont {Sunko}}, \bibinfo {author} {\bibfnamefont {J.~F.}\ \bibnamefont {Rodriguez-Nieva}}, \bibinfo {author} {\bibfnamefont {M.~S.}\ \bibnamefont {Ikeda}}, \bibinfo {author} {\bibfnamefont {T.}~\bibnamefont {Worasaran}}, \bibinfo {author} {\bibfnamefont {M.~E.}\ \bibnamefont {Sorensen}}, \bibinfo {author} {\bibfnamefont {M.~D.}\ \bibnamefont {Bachmann}}, \bibinfo {author} {\bibfnamefont {J.}~\bibnamefont {Orenstein}},\ and\ \bibinfo {author} {\bibfnamefont {I.~R.}\ \bibnamefont {Fisher}},\ }\bibfield  {title} {\bibinfo {title} {Elastocaloric signatures of symmetric and antisymmetric strain-tuning of quadrupolar and magnetic phases in \ce{DyB2C2}},\ }\href@noop {} {\bibfield  {journal} {\bibinfo  {journal} {Proceedings of the National Academy of Sciences}\ }\textbf {\bibinfo {volume} {120}},\ \bibinfo {pages} {e2302800120}
  (\bibinfo {year} {2023})}\BibitemShut {NoStop}%
\end{thebibliography}%

\section*{Methods}
\subsection{Crystal Growth}
Single crystals of \ce{PrV2Al20} were grown using an Al-based self flux method \cite{Sakai2011-ks,Tsujimoto2014-cm,VHeatCapacity}. Elemental Pr, V and Al were mixed with molar ratio 1:2:97 and loaded in an alumina crucible; the mixture was then sealed in a quartz tube filled with approximately 0.4 atm of Ar gas at room temperature. The tube was heated and held at 1150 $^\mathrm{o}$C for 6 hours, and cooled at a rate of 1$^\mathrm{o}$C/hr to 800$\sim$900 $^\mathrm{o}$C, at which temperature the tube was taken out from the furnace and centrifuged to separate single crystals from flux. The resulting crystals prefer the morphology of octahedrons with (111) facets up to lateral size of a few millimeters.
\subsection{Elastocaloric measurements}
Elastocaloric measurements at low temperatures and magnetic fields were performed in a Razorbill CS-100 uniaxial strain cell in a Quantum Design Physical Property Measurement System (PPMS). To compensate for the rapid thermal contraction of \ce{PrV2Al20} \cite{magata2016low} with respect to the Ti strain cell body, \ce{PrV2Al20} single crystals shaped into long bars along [111] were attached to mounting plates made of W with Stycast FT-2850. Temperature changes generated by the AC elastocaloric effect were probed using a \ce{RuO_x} resistive thermometer \cite{rosenberg2023nematic,ye2023elastocaloric,zic2023giant} thermally connected to the sample via a thick gold wire; the latter is attached to the sample with AngstromBond AB9110LV (Fiber Optic Center). Additional silver paint (Dupont 4922N) is applied to where the gold wire touches the sample to enhance thermal contact at the joint. The thermometer is connected to a full Wheatstone bridge with three identical thermometers thermally anchored at the bath. In the experiments, an AC strain is applied to the sample, and temperature oscillation driven by the AC strain is captured by a lock-in amplifier (Stanford Research Systems, SR860) that locks into both frequencies of the AC strain and electrical current across resistance bridge. \\

\subsection{Extracting $d^2T/\varepsilon_{111}^2$ near zero octupole points}
As discussed in the main text, to properly evaluate the octupolar susceptibility near the zero of the octupolar order parameter, $d^2T/\varepsilon_{111}^2$ is expected to be extracted at each temperature and magnetic field near the zero of $dT/d\varepsilon_{111}$. For $T\leq10$ K, this is done by fitting the observed $dT/d\varepsilon_{111}$ as a function of $\varepsilon_{111}$ with a second order polynomial, and $d^2T/d\varepsilon_{111}^2(dT/d\varepsilon_{111}\rightarrow0)$ is evaluated from the slope of the polynomial near its zero. At $T>10$ K where $dT/d\varepsilon_{111}$ appears linear with $\varepsilon_{111}$ (see Supplementary Materials), we adopt the slope from the linear fit of $dT/d\varepsilon_{111}$ with respect to $\varepsilon_{111}$ as $d^2T/\varepsilon_{111}^2$ for the final evaluation of the octupolar susceptibility.

\end{document}